\documentclass{aa}
\usepackage{graphicx}
\begin{document}
   \thesaurus{  % Section 8: Form. struct. and evolut. of stars
              (  % 08.01.1,
                % 08.06.2,
                % 11.01.1,
                % 11.05.1,
                % 11.06.1,
               )}% Stars: structure of.
%
%\received{..}
%\accepted{..}

%
\title{Models for the interpretation of CaT and the blue
 Spectral Indices in Elliptical Nuclei}

\author{M. Moll\'{a},\inst{1} \and M.L. Garc\'{\i}a-Vargas \inst{2} }

\institute{Departement de Physique, Universit\'{e} Laval,
Chemin St-Foy, G1K 7P4-Quebec, Canad\'{a}
\thanks{\emph{Present Address:} Dep. de F\'{\i}sica Te\'orica,
Universidad Aut\'onoma de Madrid, 28049 Cantoblanco, Madrid, Spain}
\and Instituto de Astrof\'{\i}sica de Canarias, GRANTECAN Project,
V\'{\i}a L\'actea S/N. 38200 La Laguna, Tenerife. Spain}

\offprints{Mercedes Moll\'{a}}
\mail{mercedes@pollux.ft.uam.es}

\date{Received xxxx 2000; accepted xxxx 2000}
  %\today

\titlerunning{Spectral indices Models for  Elliptical Nuclei}
\authorrunning{Moll\'{a} \& Garc\'{\i}a-Vargas}
\maketitle

\begin{abstract}

We present a grid of theoretical models where the calculation of
absorption line spectral indices in both the blue and red wavelength
ranges is done with the same evolutionary synthesis code.
We have computed some of these indices: CaT, Na~{\sc i},
Mg~{\sc i} in the near infrared and Mg$b$, Mg$_{2}$, Fe52, Fe53, NaD and
H$\beta$, in the blue-visible range, for Single Stellar Population
(SSP) of 6 different metallicities, (Z=0.0004, 0.001, 0.004, 0.008,
0.02 and 0.05), and ages from 4 Myr to 20 Gyr.

From the comparison of these evolutionary synthesis models with a
compilation of elliptical galaxy data from the literature, we
find that the observed CaT index follows the blue
$\langle{\rm Fe}\rangle$ index rather than Mg$_{2}$ as the models predict.
If this implies an {\sl over-abundance} [Mg/Ca] and we
take into account the masses of stars which produce Mg and Ca, these stars
could form in a time scale shorter than 5 Myr from the beginning of
the star formation process. Alternatively, an IMF biased towards very
massive stars (M$> 40 {\rm M_{\odot}}$) at the early epoch of star formation
in elliptical nuclei has to be assumed. We also suggest to
revise the calculation of the nucleosynthesis yield of Magnesium.

By using the diagnostic diagram CaT-H${\beta}$ to disentangle age and
metallicity in such populations, we obtain around solar abundances and a
sequence of ages between 4 and 16 Gyr for the galaxy sample.


\end{abstract}
\section{Introduction}

The study of chemical abundances in elliptical galaxies has
traditionally been performed through the analysis of absorption
features usually present in their spectra (see the recent review by
Henry \& Worthey \cite{HW}).  The observation of such indices in the
blue spectral range ---in particular the so called {\sl Lick}
indices--- has been a very fruitful tool to interpret the physical
properties of elliptical galaxies and globular clusters, both assumed
to consist of old stellar populations.  There are many
articles compiling observational data for some of these indices
(e.g. Trager et al. \cite{trager} and references therein), specially
in Mg$_{2}$ and $\langle {\rm Fe} \rangle =({\rm Fe5270}+{\rm
Fe5335})/2$. Some works have also measured other indices in the same
spectral region such as Mgb, NaD and H$\beta$.

Evolutionary synthesis models are the tool most frequently used to interpret
observed spectra. There are a large number of different models
(see Leitherer et al. \cite{leitherer}, and references therein ) which
have become available thanks to the development of theoretical
isochrones, computed for a wide range of ages and metallicities. An
additional basic input for these models is an atlas of stellar spectra
(empirical or theoretical) which provides the spectral energy
distribution of each elemental area of the Hertzsprung-Russell Diagram
(HRD). If the spectral resolution of the available stellar atlas is
good enough, line-strength indices can be measured directly in the
final spectrum resulting from the calculation (see Vazdekis
\cite{V99}).

When this is not the case, or when the stellar atlas consists
of atmosphere models, empirical calibrations of line-strength indices
(also known as {\sl fitting functions}) must be incorporated into
the models. These fitting functions are obtained by observing a large
sample of stars covering the widest available range of the basic
atmospheric stellar parameters (effective temperature T$_{\rm eff}$,
surface gravity $\log g$, and metallicity ---usually parameterized by
${\rm [Fe/H]}$---; some authors also include relative abundances ${\rm
[X/Fe]}$ parameters to introduce elemental ratios different from
solar). Among the most employed sets of fitting functions are those
provided by the Lick group (Gorgas et al.  \cite{Gorgas93}; Worthey et
al. \cite{WFGB94} ---hereafter WFGB94), and those of the Marseille
group (Idiart \& Freitas-Pacheco \cite{idiart95}, Borges et al.
\cite{BIFT95}; hereafter BIFT95). Examples of evolutionary synthesis
models, in which blue spectral indices for single stellar populations
(SSP) of different ages and metallicities are computed, are those of
Worthey (\cite{W94}, hereafter W94), Casuso et al. (\cite{casuso}),
Bressan et al.  (\cite{BCT96}, hereafter BCT96), Vazdekis et
al. (\cite{vazdekis96} hereafter VCPB96), and Kurth et
al. (\cite{KFF99}, hereafter KFF99). All these works have employed the
polynomial functions of WFGB94. On the other hand, BIFT95 have made use of
their own set of fitting functions, which have also been employed in
the models of Tantalo et al. (\cite{tantalo98}). Most of these
synthesis models give estimates for the blue-yellow line-strength
indices, such as Mgb, Mg$_{2}$, Fe5270 and Fe5335 (sometimes only
$\langle {\rm Fe} \rangle$), NaD and H${\beta}$.

An important result is obtained from the study of the locus of data in
the plane Mg$_{2}$--Fe (where Fe means an iron index such as Fe5270,
Fe5335 or $\langle {\rm Fe} \rangle$). The correlation followed by
globular cluster data is adequately reproduced by synthetic models of
spectral indices applied to old stellar populations of low
metallicities, a result which is not unexpected, since most of the poor-metal
stars used to calibrate the spectral index dependence on metallicity
are members of these globular clusters. This correlation is steeper
than that found for elliptical galaxy nuclei which cannot be fitted by
the models even by using the oldest and more metal-rich stellar
populations (Burstein et al. \cite{B84}; Gorgas et al.  \cite{GEA90};
Worthey et al. \cite{W92}; Davies et al.  \cite{davies93}; Carollo \&
Danzinger \cite{CD94a},\cite{CD94b}; Fisher et al.  \cite{fisher};
Vazdekis et al. \cite{vazdekis97}). In fact, elliptical galaxies are
located below the lines in the mentioned diagrams. The usual
explanation states that old elliptical galaxies formed stars very
quickly in the past, after the production of large quantities of
magnesium and other elements by massive stars (through the ejection of
metals by Type~II supernovae, SNe), but before the bulk of iron
production, which is mainly synthesized by Type~I supernovae resulting
from the evolution of low-mass stars. The iron-peak elements appear at
least 1~Gyr later than the $\alpha$-elements in the interstellar
medium. This result limits the star formation duration to less than 1
Gyr, after the start of the process

Therefore, the so-called {\sl over-abundance} of Magnesium over Iron
is actually an {\sl under-abundance} of Iron, in terms of the absolute
values of total abundances, and since Calcium is also an
$\alpha$-element, it should be expected that Ca indices follow the
Mg$_2$ behavior: if CaT and Mg$_{2}$ indices were directly related to
the abundances of Calcium and Magnesium, and both elements were mostly
produced by Type~II~SNe, one should expect that a large Mg enrichment
would also imply a large proportion of Ca in comparison with the Iron
abundance, implying [Ca/Fe] $>0$, too. On the contrary, if models are
not able to reproduce the observational data, a new explanation should
be proposed.  However, elliptical galaxies data seem to follow the
model predictions in the Ca4455--$\langle {\rm Fe} \rangle$ plane
(Worthey \cite{W98}) thus implying a $\rm [Ca/Fe]=0$.  Since this
result is not well understood, here we propose the use of the Calcium
Triplet index at $\sim \lambda8600$~\AA, CaT ($\lambda8542+8662$~\AA),
to test the predictions of theoretical models against observational
data in the plane CaT-$\langle {\rm Fe}\rangle$.  This point will be
discussed in detail in the following sections.

It is important to stress that model predictions used to compare the
variation of the CaT with other spectral features in the blue spectral
region should be obtained with the same computational technique and
inputs, i.e., the same models with identical stellar tracks and
atlases of Stellar Energy Distributions, SEDs.  For this reason, in
this paper we will present index predictions obtained with a revised
version of the evolutionary synthesis models already presented by
Garc\'{\i}a-Vargas et al. (\cite{Paper I}, hereinafter Paper I).
There we modeled the equivalent width of the two main lines of the
CaII Triplet ($\lambda\lambda 8542, 8662$~\AA), following the index
definition given by D\'{\i}az et al. (\cite{DTT}, hereafter DTT), for
SSPs with ages ranging from 1 Myr to 17 Gyr, and for 4 different
metallicities Z=0.004, 0.008, 0.02 and 0.05.  An important conclusion
derived from Paper I is that the CaT index is almost constant with
age, and only dependent on metallicity for ages older than 1 Gyr, when
the IR flux is dominated essentially by giants. This result indicates
that the CaT is a potential tool, in conjunction with other
age-sensitive indices such as H$\beta$, to confront the well-known
age-metallicity degeneracy problem in old populations. In fact, both
indices produce a quasi-orthogonal grid of constant age and
metallicity lines (see Fig.~7 in Paper I). In the revised version of
the models employed in this work, we have included the computation of
the most common indices in the blue spectral region, following the
same strategy as that employed in Paper I for the near-IR indices.

In order to check our model results about CaT in that work, we
compared the predicted indices with the globular cluster data (see
paper I), and obtained a dependence of this CaT for the oldest stellar
populations on the metallicity similar to that estimated from those
data.  Unfortunately, there were just a few CaT observations in
elliptical galaxies to compare with the model results.  In this new
piece of work we have compiled data for a sample of elliptical
galaxies, for which both the CaT and Lick indices are available in
the literature. This will allow us to compare the predictions of the
new models with the indices measured in both blue and near-IR spectral
regions.

This paper is organized in the following way: in Section~2 we give a
description of the evolutionary synthesis model, with special
attention to changes introduced with respect to Paper~I, and we
discuss the criteria followed to select the fitting functions. The
comparison of models with data is shown in Section~3. A discussion is
performed in Section~4 and finally, our conclusions are presented in
Section~5.

\section{The evolutionary synthesis model}

\subsection{Model description}

The evolutionary synthesis models presented in this paper have been
computed using the technique already described in paper~I.  The total
mass of every SSP is $1\times 10^{9}$ M$_{\odot}$ with a Salpeter-type
IMF, $\Phi(m)=m^{-\alpha}$, with $\alpha=2.35$, from m$_{\rm
low}=0.6$~M$_{\odot}$ to $m_{\rm up}=100$~M$_{\odot}$. We have
followed the passive evolution of a single--burst stellar population
(SSP) through ages from 4~Myr to 20~Gyr (with a logarithmic age step
of 0.1~dex until 10~Gyr, and 0.02~dex afterwards). This wide range is
useful to analyze composite populations, as it occurs in spiral disks
(see Moll\'{a} et al. \cite{molla}) or the non-negligible possibility
---non treated here--- of having recent star formation over-imposed
over an older stellar population (see Pellerin \& Robert,
\cite{pellerin}). In particular it might be used for studying, in
the near future, starburst galaxies where the star formation bursts were
provoked by radial flows in spiral disks which pushed the gas towards
their centers.  Thus, this work also is useful to check that this
model gives reasonable spectral indices for SSP before applying it to
other more complex stellar populations.

The spectral energy distribution for each SSP is synthesized by adding
the spectra corresponding to all the points in the theoretical HR
diagram, taken from Bressan et al. (\cite{bressan93}) . The spectrum
associated to each point was selected using the closest atmosphere
model in the stellar parameter space, and properly luminosity scaled (
see Garc\'{\i}a-Vargas et al. \cite{GVBD}).  Atmosphere models were
taken from Lejeune et al. \cite{lejeune97}, \cite{lejeune98}), who
provide an extensive and homogeneous grid of low-resolution
theoretical flux distributions for a large range in T$_{\rm eff}$
(from 2000~K to 50000~K), gravity ($-1.02 < \log g < 5.50$), and
metallicity ($ -5.0 \leq \rm [M/H] \leq +1.0$), by including M dwarf
model spectra.  We chose their corrected fluxes, color-calibrated flux
distributions, constructed through empirical effective
temperature-color relations. These spectra span from 91~\AA\ to
16000~nm, with an average spectral resolution of 20~\AA\ in the
optical range. In order to provide predictions for the youngest
stellar populations, we supplemented the above data with hot stars
(T$_{\rm eff} > 50000$~K) from Clegg \& Middlemass
(\cite{clegg}). These last stars do not exist in old stellar
populations but they will be necessary in spiral disk models, where
young stars usually contribute to the total flux.

The present models have allowed us to compute the evolution of
several line-strength indices: Na~{\sc i}, Mg~{\sc i} and CaT in the
near-IR, and Mgb, Mg$_2$, Fe5270, Fe5335, NaD and H$\beta$ in the blue
band.  The index definitions given by DTT have been used for the CaT
and Mg~{\sc i} indices, whereas the Na~{\sc i} index was computed as
in Zhou (\cite{Z91}).  The index definitions for the blue indices
(Lick/IDS system) can be found in Trager et al.  (\cite{trager}).
Although in the comparison with observational data we are only using
CaT, Mg$_{2}$, $\langle{\rm Fe}\rangle$ and H$\beta$, the calculation
of the additional indices allow to us to check whether our blue model
predictions are compatible with those obtained by other authors. In
addition, the Mg~{\sc i} and Na~{\sc i} indices in the near-infrared
are predictions which may be useful in the comparison with
high-resolution spectral data in low velocity dispersion objects.

To calculate the synthetic line-strength indices, as a function of age
and metallicity, we have followed the procedure explained in
paper~I. The Lick/IDS indices have been computed using two different
sets of fitting functions: (1) the polynomial functions from WFGB94,
and (2) the fitting functions from BIFT95.  Both sets of formulae give
the behavior of each index as a function of the main stellar
parameters: gravity, effective temperature, and metallicity or iron
abundance [Fe/H]. The BIFT95 functions allow one to include the dependence
on $\alpha$--element abundance ratios, although in this paper we have
assumed $[\alpha/{\rm Fe}]=0.0$.

Although the CaT was already synthesized in paper~I, we have decided
to recompute it here in order to consider the differences in the
$m_{\rm low}$ (we have employed 0.6~M$_{\odot}$, instead of
0.8~M$_{\odot}$), and in the atmosphere models, which now include M
dwarf spectra.  Thus, this recalculation provides us with a set of
indices obtained with the same computational technique. The only
difference between the blue and the near-IR indices presented in this
paper resides on the stellar libraries used to derive the
corresponding indices values. The inclusion of the CaT in the models
has been performed by using the predictions, based on model
atmospheres, from J{\o}rgensen et al. (\cite{JCJ92}, hereafter JCJ92).
These authors give the equivalent width of the two strongest calcium
lines as a function of the stellar effective temperature, surface
gravity, and calcium abundance, and when these functions are used for
the DTT stellar library, the predictions are in excellent agreement
with the data.  Thus, we prefer to use JCJ92 models, which revealed a
complex behavior of the calcium lines, instead of relations from DTT,
who only gave empirical relations between CaT and the stellar
parameters, as a first component analysis, in order to explain the
parametric behavior of the star sample.

\begin{figure*}
\resizebox{1.0\hsize}{!}{\includegraphics[angle=-90]{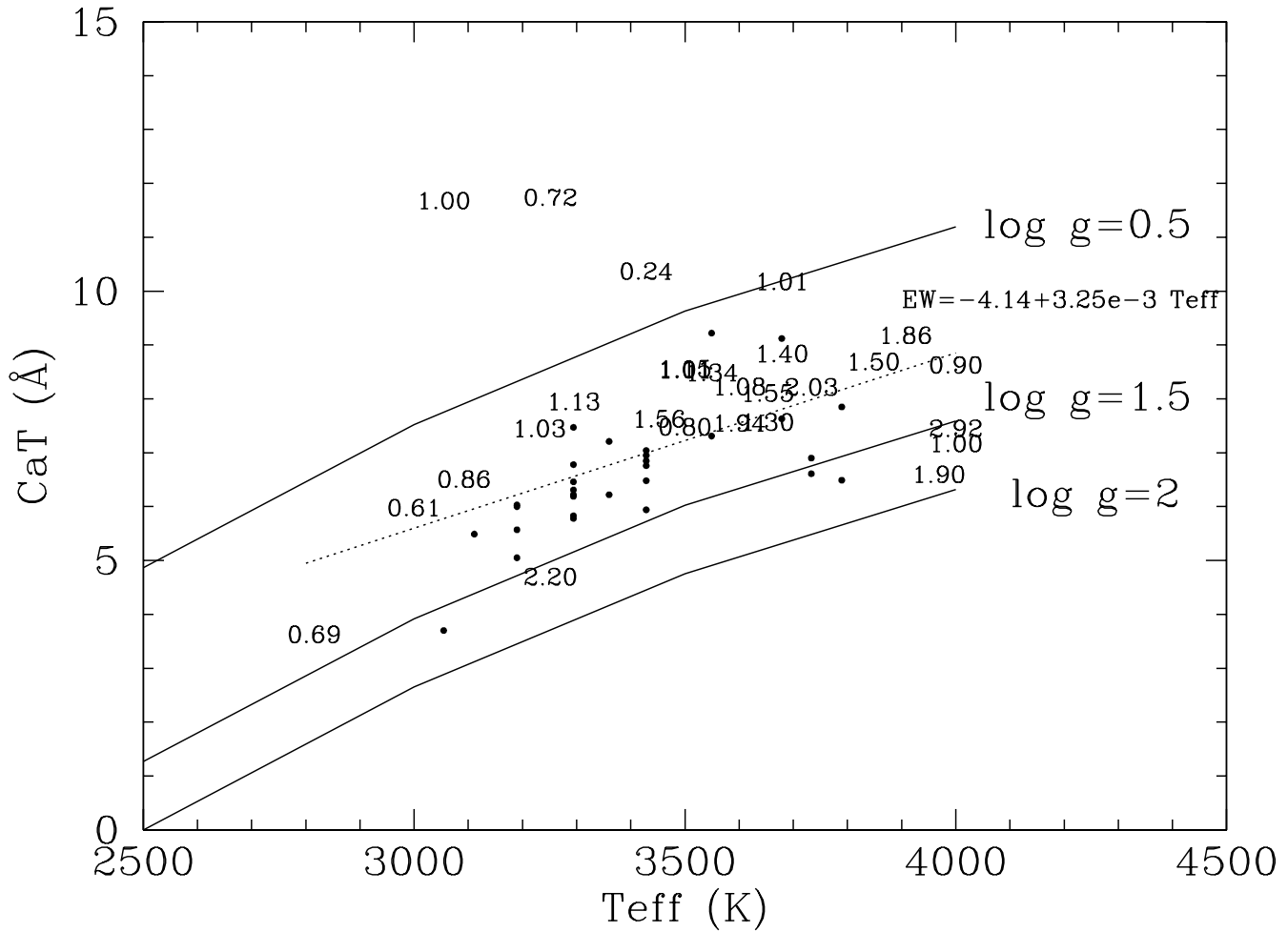}}
\caption[]{The dependence of CaT with Teff for cool stars. Solid lines
represent the extrapolation of the JCJ92 functions for three different
gravities given in the figure. Numbers are data, given with symbols
as gravity values, from Zhou (1991 \cite{Z91}) and Mallik (1997 \cite{mallik})
Solid dot are data from Zhu et al (1999 \cite{zhu}). The dotted line
is the least squares fit to all these data.}
\label{cat_teff}
\end{figure*}

It is very important to highlight that, since JCJ92 predictions are
only valid in the temperature range from 4000 to 6600~K, we have
extrapolated their fitting functions for cooler stars (see also
paper~I). Although this is always dangerous, we are confident in
the generic trend of JCJ92 results with Teff, because these relations,
extrapolated for Teff $< 4000 K$, were already compared with M stars
data in Paper I: a decreasing of CaT with Teff was found in both
cases.  To strengthen our confidence on this point, we show in
Fig.~\ref{cat_teff} the CaT-Teff relations obtained with the JCJ92
extrapolated functions for different gravities (solid lines), and
compare them to CaT data for the cool star samples from Zhou
(\cite{Z91}), Mallik (\cite{mallik}) and Zhu et al. (\cite{zhu}).  The
two first sample data are shown with symbols representing the stellar
gravity. The last set data are solid dots. The dotted line is the
least squares fit for these data, given by the equation:

\begin{equation}
CaT=3.25 10^{-3}\times Teff -4.10
\end{equation}

The dependence of the CaT on Teff has a similar slope in both,
 empirical and theoretical, cases. Data are limited by the
 extrapolation lines for $\log{g}=0.5$ and $\log{g}=2$. Those with
 lower gravities seem to be located higher in the graph than those
 with higher gravities which are lower, by following the same trend
 than theoretical models. Maybe the dependence of this theoretical
 function on gravity does not exactly reproduce the observed one for
 cool stars, but this point cannot be estimated with the small number
 of stars with known gravities used for our fit.  The influence of
 using one or the other kind of dependence (empirical equation (1) {\sl
 vs} JCJ92 functions) for cool stars on model results will be analyzed
 in the following section.  In summary, this being the main source of
 uncertainty of the present models, we attempt to use the most adequate
 solution until more confident fitting functions in this spectral
 range are available.

For Mg~{\sc i} index, the empirical calibration obtained from DTT
have been used, whereas in the case of the Na~{\sc i} index we have
followed Equations~3 and~5 from Zhou (\cite{Z91}) who calibrated
this index as a function of gravity, stellar abundance and (R-I)
Johnson-system color for spectral types from G0 to M3. For the coolest
stars we assign a value of  1\AA, following data shown by Z91
in their Figure 6. The (R-I) color  has been taken from isochrones.

\subsection{Model Results}

Table~\ref{tabla_indices_modelos} shows the line-strength predictions
for three different metallicities and six age steps, using WFGB94
fitting functions. We have selected these ages since the discussion in
this paper is focused on old populations in elliptical galaxies. (The
whole set of results for six metallicities and ages from $\log
{age}=6.60$ to $\log {age}=10.30$ is available in electronic format).
\footnote {Complete set of Table 2 is only available in electronic
form at CDS via anonymous ftp to cdsarc.u-strasbg.fr (130.79.128.5)
or via http://cdsweb.u-strasbg.fr/Abstract.html, or upon request
to authors}

\small
\begin{flushleft}
\begin{table*}
\caption[]{Spectral indices predicted with the evolutionary synthesis model
described in this work (with the WFGB94 fitting functions). We only present
results for a selected sample of ages and metallicities. The whole set of index
predictions is available in electronic format.}
\begin{tabular}{lccccccccc}
\hline
\noalign{\smallskip}
{$\log {\rm age}$} & {CaT} & {Na~{\sc i}} & {Mg~{\sc i}} &
{Mgb} & {Mg$_{2}$} & {Fe5270} & {Fe5335} &
{NaD}       & {H${\beta}$}   \\
{(yr)} & {(\AA)} & {(\AA)}       & {(\AA)} &
{(\AA)}  & {(mag)} & {(\AA)}& {(\AA)}&
{(\AA)}       & {(\AA)}   \\
\noalign{\smallskip}
\hline
\noalign{\smallskip}
\multicolumn{10}{c}{Z=0.008}\\
  9.30& 5.356& 0.025& 0.612& 2.138& 0.127& 1.937& 1.574& 1.674& 3.130 \\
  9.60& 5.614& 0.058& 0.622& 2.488& 0.145& 2.146& 1.764& 1.830& 2.461 \\
  9.90& 5.826& 0.083& 0.650& 2.907& 0.172& 2.412& 2.000& 2.013& 1.988 \\
 10.08& 5.895& 0.106& 0.650& 3.143& 0.185& 2.507& 2.082& 2.099& 1.804 \\
 10.14& 5.910& 0.111& 0.652& 3.262& 0.192& 2.563& 2.133& 2.135& 1.728 \\
 10.20& 5.922& 0.124& 0.649& 3.308& 0.195& 2.573& 2.138& 2.153& 1.687 \\
\multicolumn{10}{c}{Z=0.02}\\
  9.30& 6.815& 0.136& 0.717& 2.556& 0.163& 2.416& 2.152& 2.355& 2.795 \\
  9.60& 7.063& 0.183& 0.752& 3.335& 0.211& 2.823& 2.541& 2.747& 2.041 \\
  9.90& 7.078& 0.240& 0.759& 3.741& 0.235& 2.967& 2.688& 2.950& 1.749 \\
 10.08& 7.108& 0.255& 0.771& 4.058& 0.260& 3.148& 2.864& 3.173& 1.522 \\
 10.14& 7.071& 0.276& 0.767& 4.117& 0.267& 3.181& 2.894& 3.223& 1.468 \\
 10.20& 7.095& 0.277& 0.772& 4.203& 0.274& 3.242& 2.947& 3.294& 1.405 \\
\multicolumn{10}{c}{Z=0.05}\\
  9.30& 8.454& 0.560& 0.868& 3.393& 0.224& 3.047& 2.997& 3.406& 2.296 \\
  9.60& 8.475& 0.610& 0.888& 4.231& 0.283& 3.446& 3.408& 3.970& 1.718 \\
  9.90& 8.329& 0.717& 0.876& 4.683& 0.315& 3.627& 3.594& 4.276& 1.435 \\
 10.08& 8.285& 0.766& 0.876& 4.989& 0.342& 3.786& 3.746& 4.568& 1.282 \\
 10.14& 8.300& 0.763& 0.878& 5.081& 0.352& 3.865& 3.810& 4.710& 1.222 \\
 10.20& 8.301& 0.768& 0.878& 5.120& 0.359& 3.909& 3.856& 4.792& 1.179 \\
\hline
\noalign{\smallskip}
\end{tabular}
\label{tabla_indices_modelos}
\end{table*}
\end{flushleft}
\normalsize

The results for the CaT index are very similar to those from paper~I.
As a comparison, for an age of 13~Gyrs and abundances Z=0.004, 0.008,
0.02 and 0.05, the present models predict the following values: 4.58,
5.91, 7.08 and 8.30~\AA, respectively. The corresponding indices in
paper~I were: 4.65, 5.95, 7.32 and 9.28~\AA. These differences are
lower than 10\%, implying that the new low-mass limit does not have an
important effect on the CaT values, although we note a larger difference
(1 \AA) for Z=0.05 probably due to the effect of including M dwarf spectra
in the atlas, which is larger on the metal-rich stellar populations.

These results are also similar to those obtained by other authors, such as
 Vazdekis et al. (\cite{vazdekis96}), Idiart et
al. (\cite{idiart96}, hereafter ITFP96), Mayya(\cite{mayya}) or even
the most recent ones from Schiavon et al. (\cite{schiavon}, hereafter
SBB99).  The comparison with other models is made in
Table~\ref{comp_cat} where results for two SSP of 1 Gyr and 13 Gyr
old, with different metallicities, are shown. In all cases,
the CaT is independent of age and very dependent on metallicity,
except for the results from VCPB96 for Z$=0.05$, which are smaller
than the solar abundance ones.

\small
\begin{flushleft}
\begin{table*}
\caption{Comparison of model predictions for CaT index,
obtained by different authors and this work, for SSP's of  1 and 13 Gyr}
\begin{tabular}{lcc}
\hline
\noalign{\smallskip}
{Model} & {CaT(\AA)} & CaT(\AA)  \\
 & (1 Gyr) & (13 Gyr) \\
\hline
\noalign{\smallskip}
\multicolumn{3}{c}{Z $=$ 0.008}\\
 This work & 4.59 & 5.92 \\
 VCPB96 & 6.24 &6.87  \\
 ITFP96 &5.12& 6.25\\
 Mayya97 & 4.00 & --- \\
 SBB99& --- & 0.89 (= 6.32)\\
\multicolumn{3}{c}{Z $=$ 0.02}\\
 This work & 6.02  & 7.10 \\
 VCPB96 & 8.32& 8.21  \\
 ITFP96 &  6.20 &7.33 \\
 Mayya97 & 3.40 & --- \\
 SBB99& --- &1.00 (=7.10)\\
\multicolumn{3}{c}{Z $=$ 0.05}\\
 This work & 7.56 & 8.30 \\
 VCPB96 & 8.40 & 8.10  \\
 ITFP96 & 6.85 & 7.98\\
 Mayya97 &4.50 & ---\\
 SBB99& --- &  1.15 (=8.17)\\
\hline
\noalign{\smallskip}
\end{tabular}
\label{comp_cat}
\footnotesize \\

{Note: SBB99 do not give the total Cat in \AA but values normalized to
the 13 Gyr and solar metallicity. Values of this Table are estimated
by using our result for this age and their dependence on metallicity
and ages.}\\

\end{table*}
\end{flushleft}
\normalsize

\begin{figure*}
\resizebox{1.0\hsize}{!}{\includegraphics[angle=-90]{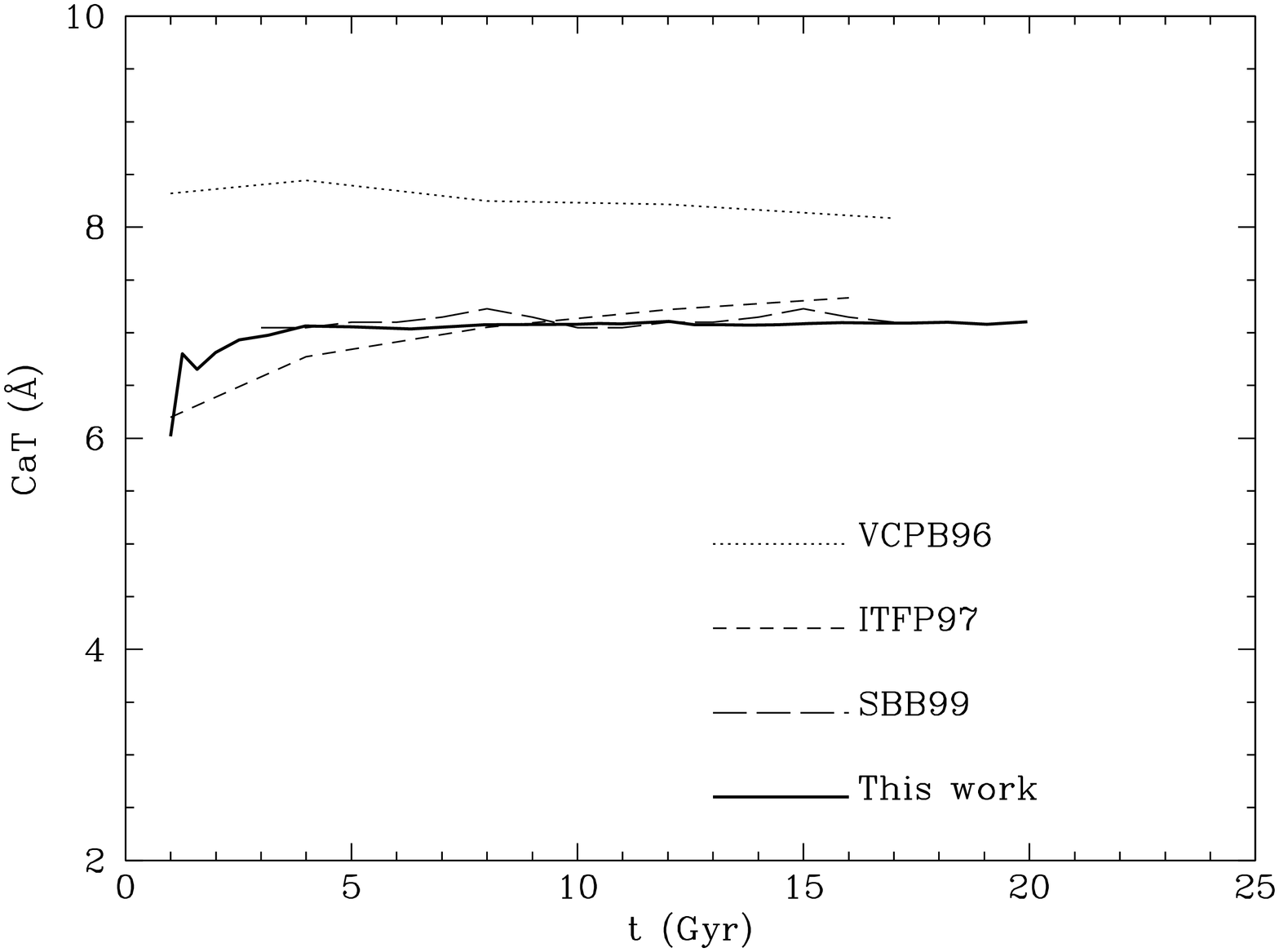}}
\caption[]{Comparison of the time evolution of the CaT index
predicted by SSP models by different authors and this work.}
\label{cat_t}
\end{figure*}

We represent the time evolution of the results of our model and
other models for
solar metallicities in Fig.~\ref{cat_t}, in order to determine whether our
conclusion about the independence of CaT on age is a general behavior
(Mayya's model is not shown because this work is only dedicated to
stellar populations younger than 1 Gyr).  We see that our model gives
an almost constant index for old stellar populations in agreement with
BCPB96 and SBB99, while models from ITFP97, who used their own stellar
library and fitting functions, give CaT values slightly increasing
with age ( In order to compare with the same kind of definitions we
have converted the CaT values from these authors to the system DTT89).
When comparing our model with BCPB96's results, (the only authors who
calculated red (CaT and Mg~{\sc i}) and blue spectral indices with the
same model) we observe that our values of CaT are ($\sim$ 1 \AA)
smaller.  This is due to the different calibration used to calculate
the stellar CaT indices: they use DTT while we use the theoretical
equations given by JCJ92. The difference between both models is almost
the same offset found between the two grids which were computed in
Paper~I with these two function sets.  Mayya (\cite{mayya}) also use
JCJ92, at least in part, reaching basically the same behavior with
the age for stellar populations younger than 1 Gyr. SBB99 also obtain
no dependence on age, although the absolute values for CaT are not
given in their work.  In conclusion, with the present knowledge about
this index we are confident in the behavior goodness for this index.

\small
\begin{flushleft}
\begin{table*}
\caption{Comparison of model predictions for the line-strength indices in the
blue spectral range obtained by different authors and this work, using solar
metallicity. Our indices are given for the two choices of fitting functions
examined.}
\begin{tabular}{lccccc}
\hline
\noalign{\smallskip}
{Model} & {Mg$_{b}$}  &
{Mg$_{2}$} & {$\langle {\rm Fe} \rangle$} &
{NaD}       & {H$_{\beta}$}\\
 & (\AA)  & (mag) & (\AA) & (\AA) & (\AA)\\
\hline
\noalign{\smallskip}
\multicolumn{6}{c}{Age $=$ 2 Gyr}\\
  W94                & 2.760 & 0.189 & 2.485 & 2.650 & 2.752 \\
  BCT96              & 2.607 & 0.168 & 2.266 & ---   & 2.904 \\
  VCPB96             & 2.056 & 0.155 & 2.108 & 2.292 & 3.440 \\
  V99                & 2.413 & 0.150 & 2.255 & ---   & 2.879 \\
  KFF99              & 2.626 & 0.177 & 2.365 & ---   & 2.722 \\
  BIFT95             & ---   & 0.089 & 2.516 & 1.272 & 2.676 \\
  This work (WFGB94) & 2.556 & 0.163 & 2.284 & 2.355 & 2.795 \\
  This work (BIFT95) & 2.017 & 0.144 & 2.236 & 1.946 & 3.136 \\
\multicolumn{6}{c}{Age $=$ 10 Gyr}\\
  W94                & 3.797 & 0.249 & 2.890 & 3.220 & 1.700 \\
  BCT96              & 3.899 & 0.255 & 2.910 & ---   & 1.540 \\
  VCPB96             & 3.921 & 0.261 & 2.906 & 3.575 & 1.491 \\
  V99                & 3.834 & 0.240 & 3.000 & ---   & 1.710 \\
  KFF99              & 3.789 & 0.253 & 2.950 & ---   & 1.590 \\
  BIFT95             & ---   & 0.196 & 2.960 & 2.460 & 1.680 \\
  This work (WFGB94) & 3.903 & 0.248 & 2.908 & 3.054 & 1.629 \\
  This work (BIFT95) & 4.098 & 0.225 & 2.927 & 2.820 & 1.606 \\
\multicolumn{6}{c}{Age $=$ 16 Gyr}\\
  W94                & 4.129 & 0.275 & 3.060 & 3.560 & 1.420 \\
  BCT96              & 4.164 & 0.274 & 3.060 & ---   & 1.350 \\
  VCPB96             & 4.188 & 0.290 & 3.141 & 4.024 & 1.197 \\
  V99                & 4.071 & 0.262 & 3.200 & ---   & 1.453 \\
  KFF99              & 4.034 & 0.274 & 3.110 & ---   & 1.360 \\
  BIFT95             & ---   & 0.244 & 3.100 & 2.980 & 1.300 \\
  This work (WFGB94) & 4.203 & 0.274 & 3.095 & 3.294 & 1.405 \\
  This work (BIFT95) & 4.856 & 0.249 & 3.113 & 3.075 & 1.398 \\
\hline
\noalign{\smallskip}
\end{tabular}
\label{tabla_otros_indices}
\end{table*}
\end{flushleft}
\normalsize

This independence of age does not depend on the extrapolation
performed for the coolest stars: when the empirical equation (1) is used,
instead of the theoretical equations from JCJ92, we also obtain a constancy of
CaT with age, although with a smaller value (6.76 \AA {\sl vs} 7.10 \AA
-- differences lower than 5 \%-- for solar abundance and $\log{age}
=10.20 $).  This behavior is explained when the isochrones information
is taken into account. When the stellar populations become older their
stars are cooler, which decreases their CaT values following
Fig.~\ref{cat_teff}.  At the same time the influence of these stars on
the total flux of the stellar population increases with age. Both
effects compensate each other, resulting in a contribution of these cool
stars to
the flux in the CaT index which remains constant with age.

\begin{figure*}
\resizebox{1.0\hsize}{!}{\includegraphics[angle=-90]{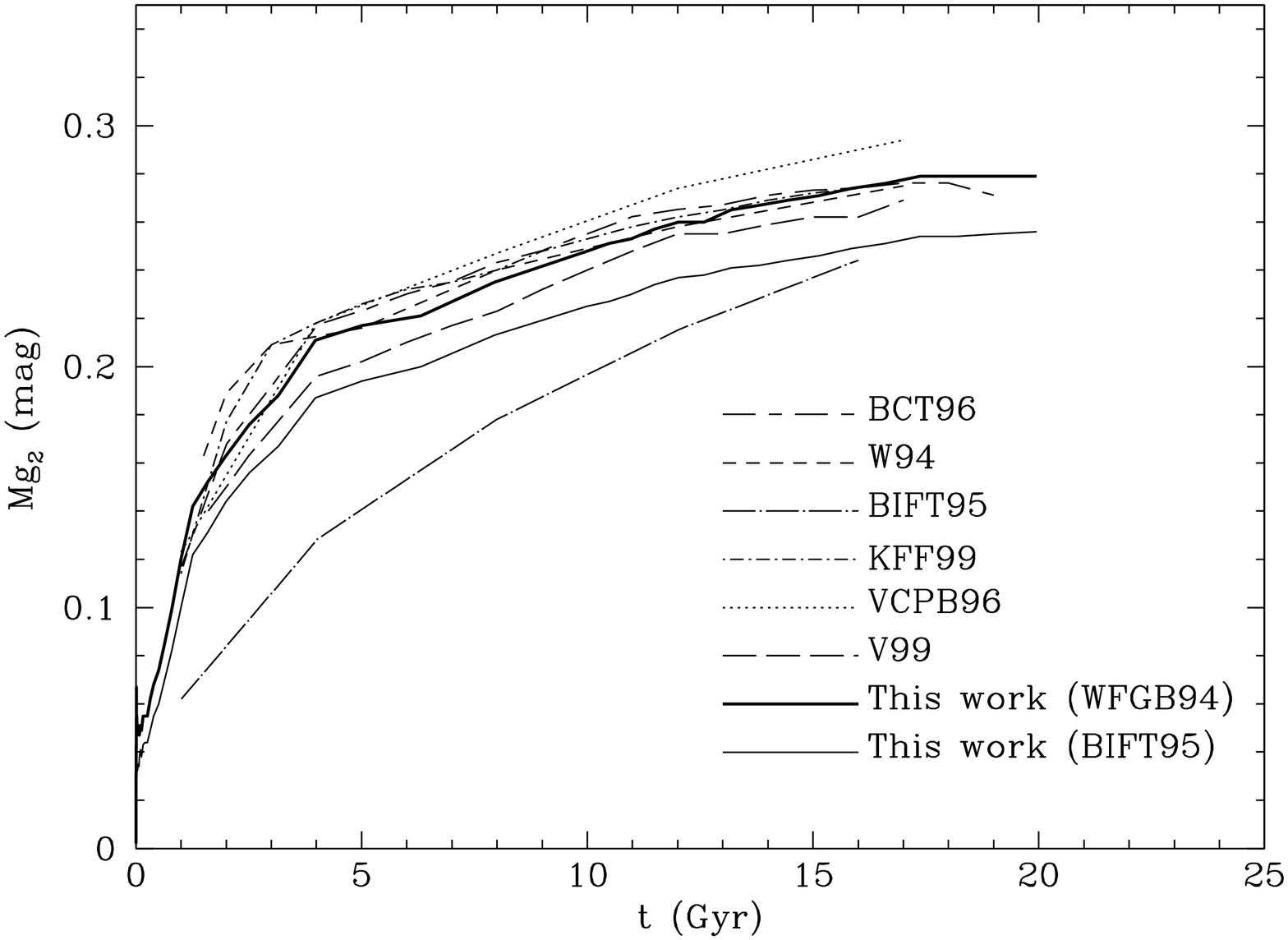}}
\caption[]{Comparison of the temporal evolution of the Mg$_{2}$ index, as
predicted by SSP models by different authors and this work.}
\label{figura_mg_t}
\end{figure*}

We have also checked our model results for the blue spectral indices
by comparing them with those derived by other authors.
Table~\ref{tabla_otros_indices} shows that our predictions are similar
to those from W94 and BCT96 when polynomial fitting functions from
WFGB94 are used. A comparison of the age evolution of the Mg$_2$ index
is shown in Fig.~\ref{figura_mg_t}.  The similarity with BCT96 is not
surprising since their input atmosphere models and isochrones are not
very different from those employed in this work. Our indices are also
close to those from KFF99 calculated with Padova isochrones, although
these authors followed a different method to assign the stellar models
in the HR diagram. VCPB96 obtain larger values of Mg$_{2}$ and lower
values of H$\beta$.  The recent estimations of V99, which are obtained
without fitting functions by measuring in the final high-resolution
synthetic spectra, are also similar to all models although slightly
lower.

The agreement between different authors is clear, with the exception of
BIFT95, which predicts lower values than all the other works. Note
that, even using the same fitting functions as in BIFT95, we are not
able to reproduce their line-strength predictions.  This not-well
understood discrepancy between the BIFT95 results and our results (and
those of other works) leads us to choose our predictions obtained
with WFGB94 fitting functions for the comparison with observational
data.

\section{Data Analysis}

\subsection{Comparison of Data with Models}

With the aim of addressing the behavior of calcium concerning the
$\alpha$-element overabundance problem, we start by comparing our
model predictions for Mg$_2$ and CaT with the available data. The set
of data is shown in Table ~\ref{tabla_elipticas}.  This index was
measured by Terlevich et al. (\cite{TDT90}, hereafter TDT90) for a
sample of galaxies, with the aim of finding a method to discriminate
between active ---i.e. with strong star formation--- and normal
galaxies. Delisle \& Hardy (\cite{DH}, hereafter DH) also estimated the CaT
equivalent width for a reduced sample of elliptical and spiral
galaxies, but they used an index definition with different
bandpasses. In our compilation there are three sources of IR data: 1)
the data from TDT90 ; 2) the set taken from DH.
These authors have kindly provided us their spectra in
order to calculate again the calcium triplet CaT with the same
definitions and windows than the first set; 3) data unpublished from
Gorgas et al. (private communication, hereafter GCGV).  All these data
 are in the same measure system.  From these sources, we have selected
 only the galaxies for which blue spectral indices are also available
 in the literature, following references of column (11). All of them
 are in the Lick system, therefore we have compiled a uniform set of
 data. In Column (12) we also show values for Mg$_{2}$ as estimated by
 Davies et al. (\cite{davies87}). These values are systematically
 lower than those more recent of Column (5).

 The measurement of equivalent widths is related to their internal
velocity dispersion through the corresponding broadening of the
spectral lines which affect both the continuum level and the
lines. The effect of the broadening is to decrease the measured EW's
(see DTT89, and their fig. 3, where this point was carefully
explained).  Therefore, the raw data have been corrected, by adding a
broadening correction to all of them, and this is how they are shown in
the figures.  These corrections are lower than the errors estimated in
most of data: for the DTT89 sample, they are $\sim 0.2 \AA$, well
within the error bars.

\small
\begin{flushleft}
\begin{table*}
\caption[]{Line-strength indices for elliptical galaxy nuclei.}
\begin{tabular}{l% Galaxy
c@{$\;\;\;$}c@{$\;\;\;$}c% CaT, error and Ref.
@{$\;\;\;\;\;\;\;\;\;$}c@{$\;\;\;$}c% Mg2 and error
@{$\;\;\;\;\;\;\;\;\;$}c@{$\;\;\;$}c% Hbeta and error
@{$\;\;\;\;\;\;\;\;\;$}c@{$\;\;\;$}c% <Fe> and error
r% reference for Mg2, Hbeta and <Fe>
@{$\;\;\;\;\;\;\;\;\;$}c@{$\;\;\;$}c% Mg2_2
}%
\noalign{\smallskip}
Galaxy &
CaT & error & Ref. &
Mg$_{2}$ & error &
H$\beta$ & error &
$\langle{\rm Fe}\rangle$ & error &
Ref. &
Mg$_{2}$ \\
 &
\multicolumn{2}{c}{(\AA)} & &
\multicolumn{2}{c}{(mag)} &
\multicolumn{2}{c}{(\AA)} &
\multicolumn{2}{c}{(\AA)} &  & (mag)  \\
\noalign{\smallskip}
\hline \noalign{\smallskip}
NGC~221  & 7.60 & 0.80 & (1) & 0.216 & 0.004 & 2.36 & 0.05 & 2.83 & 0.03 &
(5) & 0.185 \\
NGC~821  & 6.40 & 0.80 & (1) & 0.327 & 0.004 & 1.73 & 0.05 & 3.04 & 0.04 &
(5) & 0.304 \\
NGC~1052 & 8.30 & 0.80 & (1) & 0.333 & 0.007 & ---  & ---  & 2.80 & 0.15
&(11) & 0.316 \\
NGC~1700 & 6.10 & 0.80 & (1) & 0.293 & 0.004 & 2.09 & 0.05 & 3.04 & 0.04 &
(5) & 0.278 \\
NGC~2693 & 6.80 & 0.80 & (1) & 0.330 & 0.006 & 1.17 & 0.16 & 2.77 & 0.21 &
(9) & 0.328 \\
NGC~2778 & 6.89 & 0.24 & (3) & 0.349 & 0.005 & 1.76 & 0.08 & 2.88 & 0.05 &
(5) & 0.313 \\
NGC~3115 & 7.74 & 0.06 & (2) & 0.354 & 0.002  & 1.80 & 0.05  & 3.45 & 0.08
&(10) & 0.309 \\
NGC~3377 & 7.28 & 0.10 & (2) & 0.287 & 0.004 & 2.07 & 0.04 & 2.68 & 0.03 &
(5) & 0.270 \\
NGC~3379 & 7.22 & 0.14 & (3) & 0.336 & 0.004 & 1.65 & 0.04 & 2.93 & 0.03 &
(5) & 0.308 \\
NGC~3608 & 7.37 & 0.21 & (2) & 0.330 & 0.005 & 1.73 & 0.07 & 2.91 & 0.04 &
(5) & 0.312 \\
NGC~4261 & 7.28 & 0.27 & (3) & 0.351 & 0.004 & 1.43 & 0.05 & 3.17 & 0.04 &
(5) & 0.330 \\
NGC~4278 & 7.40 & 0.19 & (3) & 0.330 & 0.004 & ---  & ---  & 2.72 & 0.03 &
(5) & 0.291 \\
NGC~4431 & 6.95 & 1.13 & (3) & 0.184 & 0.005 & 1.88 & 0.20 & 2.15 & 0.14 &
(6) & ---   \\
NGC~4458 & 7.20 & 0.44 & (2) & 0.255 & 0.008 & 1.62 & 0.37 & 2.59 & 0.35  &
(8) & 0.227 \\
NGC~4472 & 7.35 & 0.20 & (3) & 0.326 & 0.006 & 1.72 & 0.07 & 3.09 & 0.06 &
(5) & 0.306 \\
NGC~4478 & 5.89 & 0.56 & (3) & 0.287 & 0.004 & 1.87 & 0.06 & 2.90 & 0.04 &
(5) & 0.253 \\
NGC~5845 & 6.66 & 0.21 & (3) & 0.316 & 0.024 & 1.63 & 0.37 & 2.88 & 0.40 &
(8) & 0.304 \\
NGC~5846 & 6.02 & 0.32 & (3) & 0.339 & 0.003 & 1.60 & 0.05 & 2.94 & 0.03 &
(5) & 0.321 \\
NGC~5846A& 7.09 & 0.40 & (3) & 0.297 & 0.002 & 1.53 & 0.08 & 2.61 & 0.05 &
(7) & 0.287 \\
\hline
\noalign{\smallskip}
\end{tabular}
\label{tabla_elipticas}

\footnotesize
\makebox[0.30\textwidth][l]{References for CaT\dotfill:}
\parbox[t]{0.7\textwidth}{
\makebox[7mm][r]{(1)}~TDT90\\
\makebox[7mm][r]{(2)}~Deslisle \& Hardy \cite{DH}\\
\makebox[7mm][r]{(3)}~Gorgas et al. (priv.comm.)\\}\\
\makebox[0.30\textwidth][l]{References for blue indices\dotfill:}
\parbox[t]{0.7\textwidth}{
\makebox[7mm][r]{(4)}~Worthey et al. \cite{W92}\\
\makebox[7mm][r]{(5)}~Gonzalez \cite{gonzalez}\\
\makebox[7mm][r]{(6)}~Gorgas et al. \cite{Gorgas97}\\
\makebox[7mm][r]{(7)}~Pedraz et al. \cite{pedraz}\\
\makebox[7mm][r]{(8)}~GEA90\\
\makebox[7mm][r]{(9)}~Trager et al. 1998\\
\makebox[7mm][r]{(10)}~Fisher et al. 1996\
\makebox[7mm][r]{(11)}~Worthey et al. \cite{W92} for Mg$_2$\\
\makebox[7mm][r]{  }~and Trager et al. 1998 for H$\beta$ and
$\langle {\rm Fe} \rangle$\\
\makebox[7mm][r]{(12)}~Davies et al. 1987 for Column (12)\\
}\\

\footnotesize
Notes:
\parbox[t]{0.8\textwidth}{
NGC~1052 and NGC~4278 present H$\beta$ in emission \\
NGC~2778 and NGC~4261 present H$\beta$ with emission contamination\\}\\

\end{table*}
\end{flushleft}
\normalsize

We have already discussed that if magnesium and calcium were produced
by the same type of massive stars, they should also trace the same
mean abundance, in which case the [$\alpha$/Fe] overabundance found in E
galaxies when studying the Mg lines should also be present in the
analysis of the Ca features. We would expect to find all the
observational points within the model lines.

\begin{figure*}
\resizebox{1.0\hsize}{!}{\includegraphics[angle=-90]{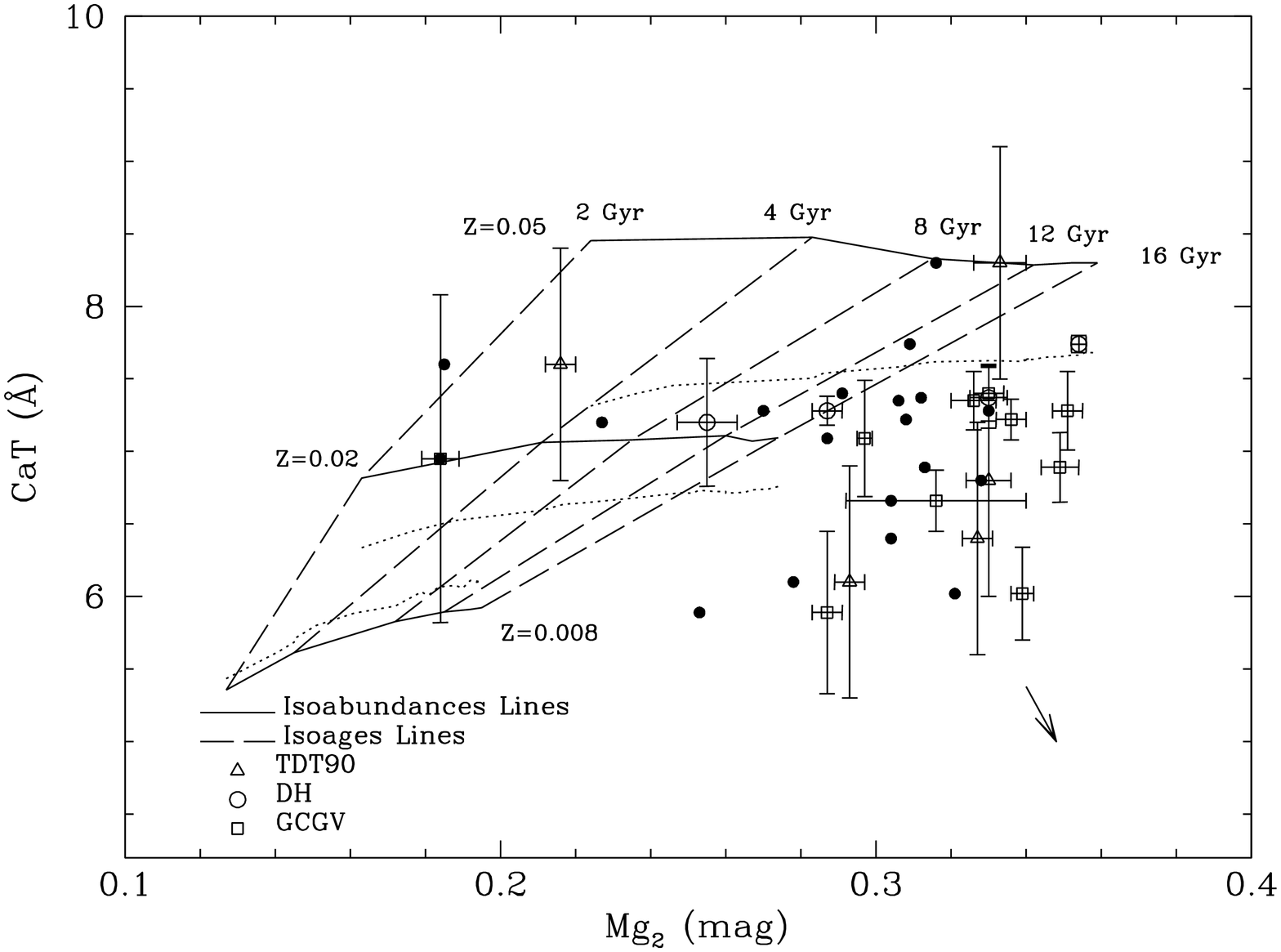}}
\caption{The index CaT {\sl vs} Mg$_2$. Solid lines and long-dashed
lines join isoabundance and isoage stellar populations, respectively,
as explained in the inset. Dotted lines represent results for
isoabundance stellar populations obtained by using the empirical
equation (1), instead of the extrapolated theoretical equations
JCJ92. Symbols refer to the observational data presented in
Table~\ref{tabla_elipticas}: open symbols with error bars
refer to Mg2 from Column (5), while full symbols
represent values from Column (12). The small vector on the figure
shows the shift of model results which would appear if the effective
temperature of the RSG branch was decreased by 200 K.}
\label{cat_mg}
\end{figure*}

This comparison is graphically shown in Fig.~\ref{cat_mg}, where the
lines correspond to the model results presented in
Table~\ref{tabla_indices_modelos}: solid lines are isoabundance
predictions (Z=0.05, 0.02 and 0.008, from top to bottom), whereas
dashed lines indicate isoages values (ranging from~2 to 16~Gyr, from
left to right). Interestingly, model predictions in this diagram
indicate that Mg$_2$ and CaT do not exhibit a strong age--metallicity
degeneracy, which means that this index--index plane could be useful
for the overabundance study. It is clear from the figure that, for
single populations older than 2~Gyr, the CaT index is roughly constant
for a given abundance, as we already explained in the previous
section, whereas Mg$_2$ is sensitive to both age and metallicity.

When the data compiled in Table~\ref{tabla_elipticas} are plotted
in this diagram (using different symbols to distinguish distinct CaT
sources, as explained in the plot key), it is apparent that the
observed Mg$_2$ indices spread beyond the model lines. This effect is
equivalent to that found by Worthey et al. (\cite{W92}) in the
Mg$_{2}$--$\langle {\rm Fe} \rangle$ plane.  In this figure, we use
solid dots to represent the values of Column (12), while the open
symbols are those of Column (5). Both kinds of points represent the
same galaxies with different estimations of Mg$_{2}$.  It is clear
that an uncertainty range appears to be involved in the calculation of the
index: there exists an offset between Davies et al. (\cite{davies87})
and data from other authors, mostly Gonzalez (\cite{gonzalez}), models
being closer to the observations when the first set is used.

The observed CaT values indicate that the calcium abundance is about
solar. Note that the location of data is not dependent on the CaT
source (although TDT90 indices show a large abundance scatter, ranging
from Z=0.02 to Z=0.05, but some of them are active galaxies).

\begin{figure*}
\resizebox{1.0\hsize}{!}{\includegraphics[angle=-90]{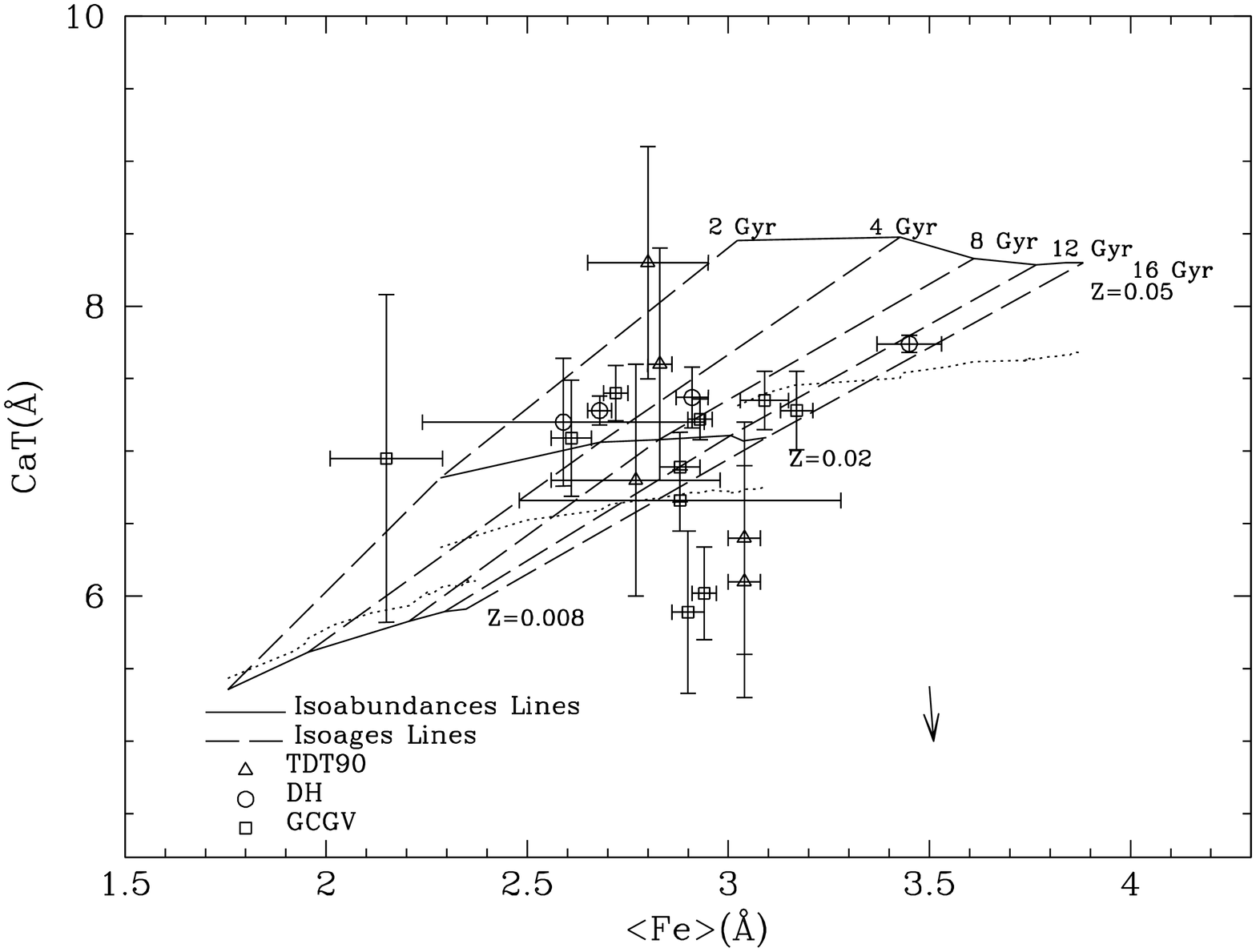}}
\caption{The index CaT vs. $\langle {\rm Fe} \rangle$, both in \AA.
The lines  and symbols
have the same meanings as in previous figures.}
\label{cat_fe}
\end{figure*}

Another piece of information came from the study of the
CaT--$\langle{\rm Fe}\rangle$ plane, shown in Fig.~\ref{cat_fe}. Model
predictions exhibit a larger degeneracy than that observed in
Fig.~\ref{cat_mg}.  It is important to note, however, that CaT and
$\langle{\rm Fe}\rangle$ are not completely degenerate due to the
dependence of the iron indices on age, almost negligible for the
CaT. This result can be easily quantified using the {\sl metal
sensitivity\/} parameter ($\Delta\log{\rm age}/\Delta\log{\rm Z}$)
defined by Worthey (\cite{WFGB94}): for CaT is 134, while for Mg2 this
value is $\sim 1.8$ and for Fe5270 is $\sim$ 2.3. It is clear from
this calculation that the CaT is an excellent abundance indicator,
highly surpassing Fe5709, the iron index most sensitive  to metallicity
($\Delta\log{\rm age}/\Delta\log{\rm Z}=6.5$) in the optical range.

When the observational data set is included in the above plane (see
Fig.~\ref{cat_fe}), it is clear that model predictions reproduce the
location of most elliptical galaxy nuclei. In addition the abundances
read from these models are mostly solar, for ages between 4--16 Gyr
(note that this result is logically the same derived from
Fig.~\ref{cat_mg} since metallicity is derived from the same index,
i.e. CaT).  There are also some outliers, but most of them correspond
to data with large error bars in CaT and/or in
$\langle{\rm Fe}\rangle$.  It is clear that most elliptical galaxies
fall within the model predictions grid.

This diagram confirms the previous trend obtained when comparing
$\langle {\rm Fe} \rangle$, with Ca4455 (e.g. Worthey \cite{W98}). We
conclude then that the simultaneous analysis of Figs.~\ref{cat_mg}
and~\ref{cat_fe} adds further weight to the idea that calcium follows
iron instead of magnesium, reinforcing the observational evidence that
calcium and magnesium behave differently in elliptical galaxies (see
section \ref{discus} for a discussion).

In order to check the possible influence of the extrapolation
used on this diagram, we have also shown, as dotted lines, in
Fig.~\ref{cat_mg} and Fig.~\ref{cat_fe} the results obtained when the
empirical equation (1) is used for the stars cooler than 4000.The CaT
values are lower but our former conclusions hold: there are a
large number of points falling out of the diagram in Fig.~\ref{cat_mg}
while in Fig.~\ref{cat_fe} some data now fall out but also in the
opposite direction.

Other possible error source comes from the uncertainties in the RGB
temperature which may be 200 degrees cooler or hotter. We have
estimated the effect of a reduction of 200 K in the effective
temperature for stars in the RGB on our indices predictions, by
indicating these variations in each figure as an arrow located in a
corner of  each diagram. This shift is not sufficient to reach the data
region in Fig.~\ref{cat_mg}, even using the empirical calibration for
the coolest stars.

\subsection{Disentangling age and metallicity: the CaT-H$\beta$ plane}

One of the main problems to understand stellar populations in
early--type galaxies is how to disentangle age and metallicity
effects. Gonzalez (\cite{gonzalez}) showed that the combination of the
H$\beta$ line-strength with indices like Mg$_2$ or $\langle {\rm Fe}
\rangle$ could break the degeneracy. It is well known that H$\beta$ is
affected by nebular emission. To overcome this problem, higher-order
Balmer lines, like H$\gamma$, have been proposed as a powerful
alternative (Jones and Worthey \cite{Jones95}; Vazdekis and Arimoto
\cite{V99b}). Unfortunately, not many accurate data on this feature
have been presented in previous works (but see Kuntschner and Davies
\cite{KD}). Since the aim of this work is not to derive absolute ages
and metallicities but to gather information from the relative trends
in the index--index diagrams, we have preferred to use the classical
H$\beta$ index, although we note that some particular galaxies may be
partially affected by an emission component (see notes in
Table~\ref{tabla_elipticas}).

\begin{figure*}
\resizebox{1.0\hsize}{!}{\includegraphics[angle=-90]{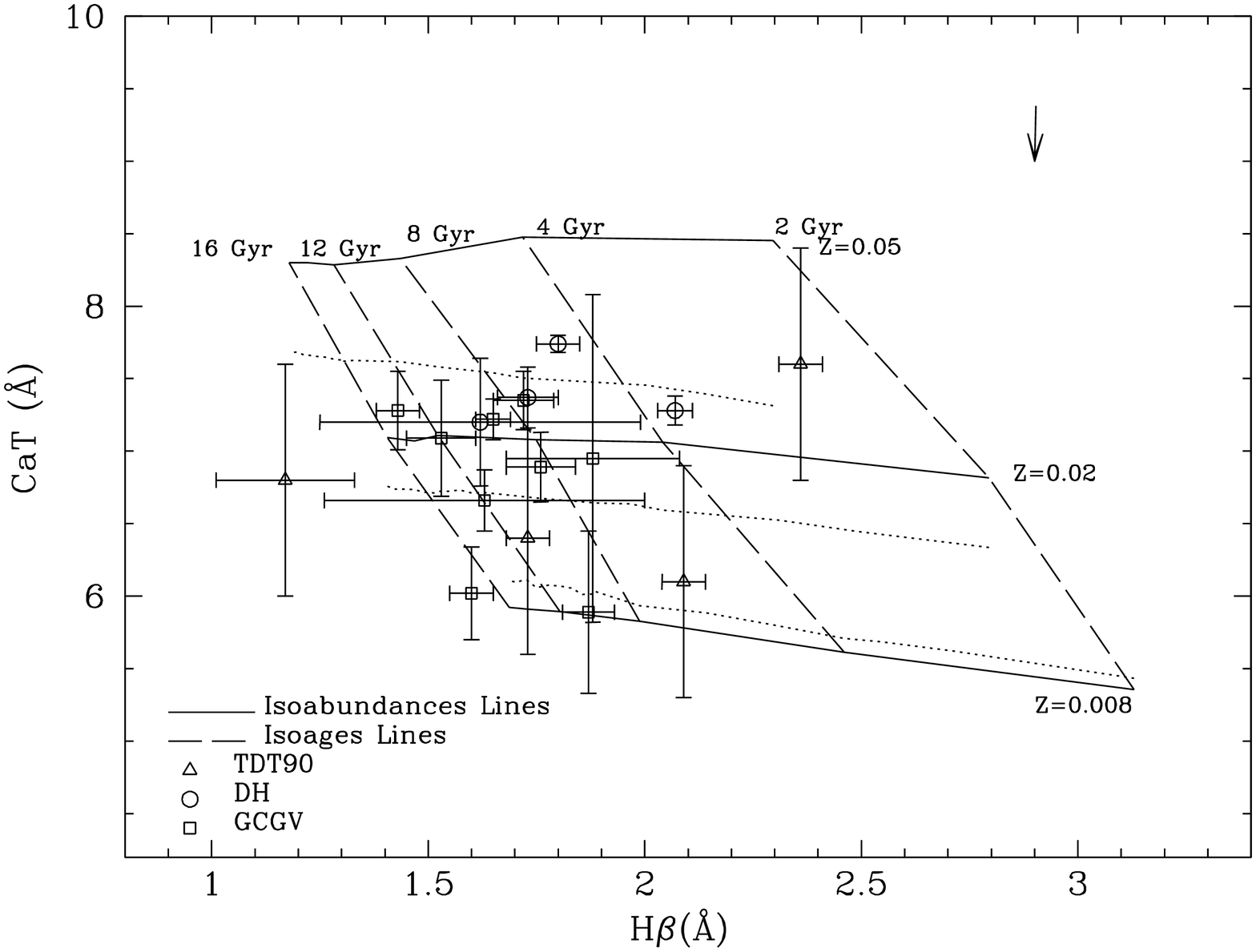}}
\caption{The diagram CaT-$H\beta$. The lines  and symbols
have the same meanings as in previous figures.}
\label{cat_hbe}
\end{figure*}

The new model predictions in the H$\beta$--CaT plane are displayed in
Fig.~\ref{cat_hbe}. The simultaneous computation of blue and near-IR
indices performed in this work allows us to confirm the previous
finding of Paper I, already discussed in the introduction to this
paper, concerning the orthogonality of the CaT--H$\beta$ diagram and
the advantage of this diagram to separate age and metallicity effects.
Almost all the data points fall inside the new model predictions,
within their error bars. Once again we represent the results
obtained through the empirical fit for CaT as dotted lines and an arrow
is included in the graph to indicate the possible shift of points if
the RGB effective temperature is reduce 200 K.

Taking the models literally, the above diagram indicates that the
stellar populations exhibit roughly solar abundances for [Ca/H].  This
is naturally the same result obtained from Figs.~\ref{cat_mg}
and~\ref{cat_fe}. In addition, the derived ages range from 4 to
16~Gyr, except for NGC~221, with an age of $\sim 2$~Gyr, which
is not too far from the result derived by Vazdekis \& Arimoto
(\cite{V99b}), who found that the better fit of a synthetic spectrum
to this galaxy is obtained for ages in the range from 2.5 to 5~Gyr.
Obviously, these results are not new since they rely on the H$\beta$
indices.

An important problem associated with the use of the Balmer absorption
features is that their continuum bandpasses usually include metallic
lines.  Another important drawback, specially in the case of H$\beta$,
is that it is well established that a large fraction ($\la 50$ \%) of
early-type galaxies exhibit Balmer emission lines at some extent.
Therefore, the determination of ages is not totally safe.

\section{Discussion}
\label{discus}

As we have already mentioned, a ratio \mbox{$[{\rm Mg}/{\rm Fe}]>0$}
can be explained by taking into account that both elements are
synthesized in the interior of stars covering a different range of
stellar masses. Invoking the same reasoning for calcium, it can be
concluded that this element is generated in another type of stars than
those synthesizing Mg. In this sense, Worthey (\cite{W98}) has
suggested that Type~II SNe could have two {\it flavors\/}, just like
Type~I and~II SNe produce iron and magnesium at different times in the
evolution of a galaxy.

It is important to note that the ratio [$\alpha$/Fe] is high because
the iron abundance [Fe/H] is low, i.e., an iron deficiency translates
into an over-abundance for alpha-elements. In particular, if one
assumes [$\alpha$/Fe]=+0.4 for a given total abundance Z, it means
that, for the total abundance Z, $\log{Z/Z_{\odot}}$ is approximately
the same and [Fe/H] is $\sim -0.4$~dex smaller.  Isochrones do not
change very much when alpha-elements are enhanced because abundances
of the principal elements are also enhanced; but the iron abundance is
decreased (see Tantalo et al. \cite{tantalo98}).

The overabundance in $\alpha$-elements, or underabundance in [Fe/H],
which can be seen in the models, occurs because the bulk of
the stars are created before the iron is produced. It may
be seen when the star formation is low, usually in low-metallicities
regions. Thus, the ratio [Ca/Fe] is over-solar in metal-poor stars
(Wallerstein, \cite{wallerstein}; Hartmann \& Gehren, \cite{Hartmann};
Zhao \& Magain, \cite{zhao}; Gratton \& Sneden, \cite{gratton}), such
as the [Mg/Fe] is. This overabundance decreases until solar ratios
are reached when the iron abundance increases.
For disk dwarfs, Edvardsson et
al. (\cite{edvardsson}).  determined the calcium abundances in the
range \mbox{$-1.0 < {\rm [Fe/H]} < 0.2$~dex}, finding that
\mbox{[Ca/Fe] $\sim 0.25$~dex} at [Fe/H]$=-1.0$, and then decreases to
solar values just following the magnesium. This kind of behavior may
also occurs for high metallicities if the time scale for the formation
of the bulk of stars is short enough to create stars before the iron
appearance, $\sim 1$~Gyr.  McWilliam \& Rich (\cite{mcwilliam})
showed, for the Galactic bulge stars, an overabundance in [Ca/Fe]$\sim
0.3$, slightly lower than those of magnesium and silicon.

Chemical evolution models, including recent theoretical
calculations of yields (Timmes et al. \cite{timmes}),
reproduce the Galactic abundances for the so called intermediate
alpha-elements (Mg, Si, Ca and Na) as the result of the evolution of
the massive stars. (We must note that it is necessary to increase the
magnesium yield by a factor of 2 in order to obtain the solar
abundance of Mg). Thus the bulk of the star formation should take a
time which is shorter than $\sim 1$~Gyr (the minimum time that has to
elapse before the low-mass stars eject the iron in the Type~I SNe
explosions) in regions where [$\alpha/{\rm Fe}]>0.0$, and longer for
the systems where [Mg/Fe]=0.0.  This result is well reproduced by
chemical evolution models (after adequately fitting the data in the solar
neighborhood): in Fig.~5 of Molla \& Ferrini (\cite{MF}) we can see a
similar behavior for Mg, Ca and Si when stars are formed in a
short time scale ($\sim 0.8$ Gyr).

\begin{figure*}
\resizebox{\hsize}{!}{\includegraphics[angle=-90]{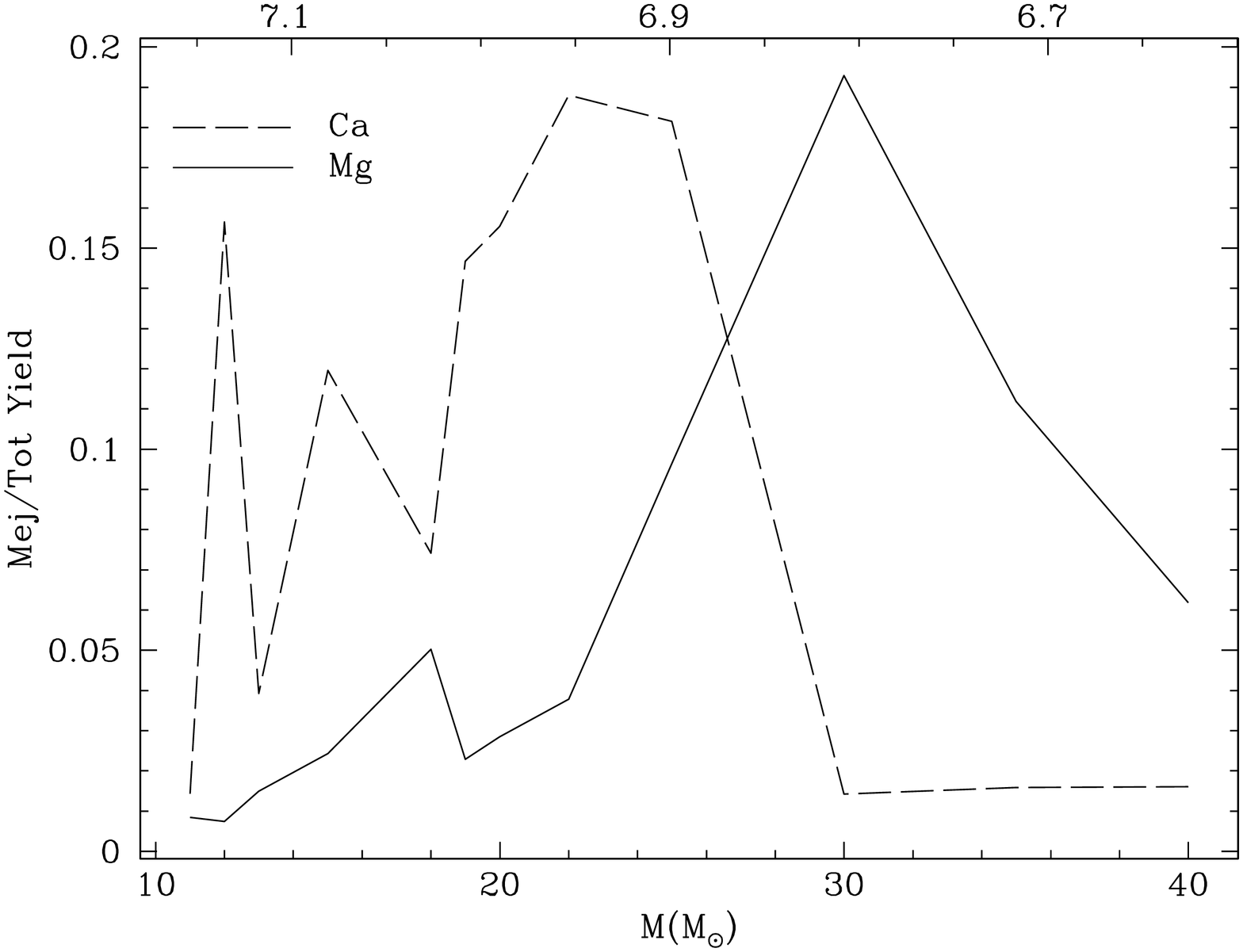}}
\caption{Magnesium and calcium yields taken from Woosley \& Weaver (1995)
as a function of the stellar mass (lower
x-axis) and mean stellar life (upper x-axis, years in logarithmic scale).}
\label{yields}
\end{figure*}

The fact that ${\rm [Mg/Fe]} > 0 $ in elliptical galaxy nuclei
implies that star formation may extend over a time shorter than 1~Gyr. If
we now introduce the observed [Ca/Fe] ratio into this reasoning, and
we assume that both calcium and magnesium are produced by similar
Type~II SNe (although with different masses), we can constrain even
more the duration of the star formation episode in these regions. In
Fig.~\ref{yields} we represent, simultaneously, the fraction of
ejected mass of $^{24}$Mg and $^{40}$Ca over the total yield for each
element, taken from Woosley \& Weaver (\cite{WW95}), as a function of
the stellar mass (lower x-axis) and the mean stellar life (upper
x-axis, in logarithmic scale). It is clear from this figure that
calcium is produced by stars with mass in the range
12--30~M$_{\odot}$, while magnesium is generated in stars of
20--40~M$_{\odot}$. This is in agreement with the more recent
metallicity-dependent yields from Portinari et al.
(\cite{portinari}), who show in their Fig.~4 the ejecta of
each element as a function of the CO-core, indicating that calcium
proceeds from lower mass stars than those which produce
magnesium. Considering the lifetimes of the stars responsible for the
bulk production of Ca and Mg, we find these values peak at $\sim 8.9$
and $\sim 6.3$~Myr, respectively.  Therefore, even with a short
difference between both lifetimes ($\sim 2.6$~Myr), there
exists a time period at which the magnesium has already been ejected
by the most massive stars but the bulk of the calcium has not yet been
released.  After this small delay, the calcium appears in the ISM, and
stars formed afterwards will incorporate both elements, Mg and Ca,
showing [Mg/Fe] and [Ca/Fe] larger than zero and a ratio [Mg/Ca]
tending to solar.

The consequence of this reasoning is important: if in elliptical
galaxy nuclei calcium is not over-abundant, while magnesium is, their
stellar populations could have been formed, at least {\it locally\/},
in a very short burst, lasting a timescale smaller than a few Myr.
\footnote{A word of caution must be said when considering the speed of
 the star formation in a stellar system. The quoted times always refer
 to {\it local\/} times, i.e.  those needed to produce the bulk of the
 star formation locally, although the required times to form the whole
 system, for instance a galaxy, could be much larger.} If
the star formation took more than this time, the calcium would
have been incorporated into the new stars, and one should expect to
observe ${\rm [Mg/Ca]} \simeq 0$, which does not seem to be the case.
This is in agreement with the dissipative theory of galaxy formation
where the collapse time scale must be shorter for more massive
galaxies by producing a correlation between the time scale of the star
formation in these galaxies and the total mass. This is supported by
Richer et al. (\cite{richer}), who claim that [O/H]
varies systematically with the velocity dispersion for spheroidals,
bulges and ellipticals, by relating the gravitational potential well
with this effect and by concluding that the time scale governs [O/Fe]
or [Mg/Fe]. The result implies a short time scale for star formation
in more massive elliptical and bulges, supporting the old idea that
ellipticals are simpler than low mass galaxies from their star
formation history.

However, the deduced constraint for the local star formation time might
seem to be too short. Then a new scenario, which could include the two
Type~II SNe {\it flavors\/}, must be invoked.  This effect can also be
enhanced (or even dominated) by a flatter initial mass function due to
special conditions (dense environment or high metallicity). If an IMF
is biased towards massive stars, the magnesium produced or {\sl yield}
in these elliptical nuclei will be larger, allowing one to find
overabundances of magnesium and also of other elements ejected by the
most massive stars such as oxygen, but not of those ejected by less
massive stars.  An IMF universal and constant is still a matter of
discussion (Larson \cite{larson}, Padoan et al.  \cite{padoan97},
Meyer et al. \cite{meyer}, Chiosi et al. \cite{chiosi}), but some
recent works claim that it must be  constant (Wyse \cite{wyse},
Tsujimoto et al.  \cite{tsujimoto}, Chiappini et al. \cite{CMP}).

One point that may enlighten this problem proceeds from the
nucleosynthetic yield calculations: by using the existing yields in a
chemical evolution model, it is not possible to reproduce the solar
abundance [Mg/H]. It is necessary to add magnesium to the stellar
production to reach the estimated solar value.  The quantity of
ejected magnesium must be a factor of $\sim$2 greater that the present yield,
implying that Fig.~\ref{yields} must change. If the deficiency in
magnesium might be accounted for by the production of the most
massive stars (M$> 40 \rm {M_{\odot}}$), we might solve two problems:
chemical evolution models would reproduce the solar abundances and the
difference between stellar lifetimes for stars producing Mg and Ca
would be large enough to allow the formation of stellar populations
with an overabundance [Mg/Ca] in a sufficient time.

Another possible solution to this problem, besides the actual ratio
[Mg/Fe] non solar, may be related with the measures of Mg$_{2}$ as
we see by comparing columns (12) and (5) from
Table~\ref{tabla_elipticas}. For some galaxies there exist data from
different authors and there are differences as large as 0.05, that is,
15 \%. Following Goudfrooij \& Emsellem (\cite{Goudfrooij}) the
possible emission lines may affect the measures of absorption
lines.  They estimated that the index Mgb may be artificially
enhanced by 0.4-0.1 \AA\ and Mg$_{2}$ by 0.03 mag, due to the [N~{\sc
i}] emission-line doublet at 5199 \AA. Taking into account that 50 \%
of giant elliptical galaxies exhibit H${\alpha} +$ [N~{\sc ii}]
emission, maybe the actual Mg$_{2}$ values must be reduced. If we
reduce the Mg$_{2}$ data by 15 \%, most of them might be almost
reproduced by models, as we show in Fig.~\ref{cat_mg} where the two
sets of data (open {\sl vs} full) are represented.

In this context, a surprising result has been obtained by Origlia et
al. (\cite{origlia}).  These authors have measured the strength of the
infrared absorption line at 1.59~$\mu$m, which is primarily sensitive
to the total silicon abundance.  These authors used this absorption to
estimate the ratio [Si/Fe] necessary to reproduce their EW, by
comparing a sample of elliptical galaxies and globular clusters with
models. They found that silicon could be enhanced by about 0.5~dex in
both kinds of stellar systems. Since Si and Ca proceed from stars with
similar masses (Portinari et al.\cite{portinari}), both should be
overabundant at the same time, contrary to our and their findings and
to our explanations for the existence of an overabundance in [Mg/Ca].
Thus, the question remains unclear until more precise calculations of
nucleosynthesis yields and observations of alpha-elements become
available.

\section{Conclusions}

We have developed an evolutionary synthesis model with which we have
produced a grid of models for SSP at 6 metallicities and a wide age
range. This code is able to predict indices in the blue-visible
spectral range, the classical {\sl Lick} indices and indices in the
near-IR such CaT, Mg~{\sc i} and Na~{\sc i} at the same time.

We have carefully analyzed the behavior of this index for the
coolest stars (Teff $<$ 4000), given by some samples of data available
in the literature, obtaining the generic trend of CaT decreasing
with effective temperature, in agreement with the extrapolated JCJ92's
theoretical functions. Therefore, although systematic effects may
still be present in the theoretical predictions, mainly due to
potential errors in the extrapolation of these JCJ92 equations, we use
the most adequate solution until more reliable fitting functions become
available.

We have compiled a set of data from the literature for galaxies for
which both kind of indices, blue-visible and near-IR, had been
observed, and we have compared their predictions with data.

 We have used our results to study the relationship between different
indices for old stellar populations by producing diagnostic diagrams
in which observed data and models can be plotted to determine the
basic physical properties of the dominant stellar population in these
galaxies.  We find that most of the galaxies with known data for Mg2, CaT
and $\langle {\rm Fe} \rangle$ remain in the CaT -- Mg$_{2}$ plane at
the place of Z=0.02, while they seem to have overabundances of Mg. This
conclusion is in agreement with other data of Ca4300 in the blue
region found by Worthey (\cite{W98}), but raises the question of
how it is possible to have solar metallicity ratio for the calcium
element and over-solar ratio for Mg abundance, while both are
$\alpha$-element produced by the same type of massive stars.

If we accept the assumption of a relative abundance [Mg/X]$>0$,
adopted to explain this kind of diagram and that it is due to a short
time scale for the star formation, and we apply the same argument for
the [Mg/Ca], this would imply that the star formation time scale in
elliptical galaxy nuclei must be shorter than $\sim 5-10$ ~Myr.
 Otherwise, we should not find a discrepancy between the data and the
models in the later diagnostic, where both indices proceed from the
same kind of alpha-elements, which is not the case. We suggest that an
update of the nucleosynthesis yields of Mg, increasing the production
of Mg for the more massive stars, may solve this problem, by extending
the elapsed time between the production of Mg and of Ca and, in
consequence, the time scale for the star formation.  An alternative
explanation might be an IMF biased towards the massive stars ($\rm M>
40 M_{\odot}$) in the early phases of star formation in elliptical
galaxies.

We must keep in mind that the emission over these spectra also may affect
Mg$_{2}$ data: they may be reduced by 5 \% if a careful analysis is
done before obtaining the spectral indices Mgb and Mg2.

We use the orthogonal diagram CaT -- H${\beta}$ to date elliptical
galaxies and to determinate their abundances, reaching the
conclusion that elliptical galaxies are nearby solar in their
abundances of calcium, in the same way as for iron, and that their ages
range between 8 -16 Gyr.

A large campaign of observations in the near-infrared to estimate the
CaT index, {\sl e.g.}  in the same set of galaxies given in Davies et
al. (1987), would be very useful to date them and to determinate their
metallicities/abundances in a clear way.

\begin{acknowledgements}

We thank J. Gorgas and N. Cardiel for the permission to use their
CaT data before publication and for fruitful discussions.  We
are grateful to Sonya Delisle and Eduardo Hardy who kindly sent us
their spectra in the red bandpass. We thank the referee, Guy Worthey,
for his useful comments for the improvement of this paper.  We also
thank A. I. D\'{\i}az for suggestions in the final version of this
work and J. Gea Banacloche for the help in the english version.
  M.M. thanks the Universit\'{e} Laval (Quebec) for the nice period,
during which part of this work was done.  M.M. has been supported
by a post-doctoral fellow of the Spanish {\sl Ministerio de Educaci\'{o}n y
Cultura}.  This work has made use of the NASA
Astrophysics Data System.
\end{acknowledgements}

\end{document}